\documentclass[reprint,amsmath,amssymb,aps]{revtex4-2}

\usepackage{graphicx}
\usepackage{dcolumn}
\usepackage{bm}

\begin{document}

\preprint{APS/123-QED}

\title{Search for the $f(R,T)$ gravity functional form via gaussian processes}

\author{J.A.S. Fortunato}
 \email{jeferson.fortunato@edu.ufes.br}
\affiliation{PPGCosmo, Universidade Federal do Esp\'irito Santo (UFES), Centro de Ci\^encias Exatas, Departamento de F\'isica - Avenida Fernando Ferrari 514, 29075-910, Vit\'oria, ES, Brazil}
\author{P.H.R.S. Moraes}
 \email{moraes.phrs@gmail.com}
\affiliation{Centro de Ci\^encias Naturais e Humanas (CCNH), Universidade Federal do ABC (UFABC) - Avenida dos Estados 5001, 09210-580, Santo Andr\'e, SP, Brazil}
\affiliation{Laborat\'orio de F\'isica Te\'orica e Computacional (LFTC),
 Universidade Cidade de S\~ao Paulo (UNICID) - Rua Galv\~ao Bueno 868, 01506-000 S\~ao Paulo, Brazil}
\author{J.G. de Lima J\'unior}
\email{grimario.lima@fis.ufal.br}
\affiliation{Instituto de F\'isica, Universidade Federal do Alagoas (UFAL) - Avenida Lourival Melo Mota S/N, 57072-970, Macei\'o, AL, Brazil}
\author{E. Brito}
\email{eliasbaj@ufob.edu.br}
\affiliation{Centro de Ci\^encias Exatas e das Tecnologias, Universidade Federal do Oeste da Bahia - Rua Bertioga 892, 47810-059, Barreiras, BA, Brazil}
%\author{T.S. Guerini}
%\address{Instituto de F\'isica, Universidade Estadual do Rio de Janeiro (UERJ) - Rua S\~ao Francisco Xavier, 524, 20550-013, Rio de Janeiro, RJ, Brazil}

\begin{abstract}

The $f(R,T)$ gravity models, for which $R$ is the Ricci scalar and $T$ is the trace of the energy-momentum tensor, elevate the degrees of freedom of the renowned $f(R)$ theories, by making the Einstein field equations of the theory to also depend on $T$. While such a dependence can be motivated by quantum effects, the existence of imperfect or extra fluids, or even a cosmological ``constant'' which effectively depends on $T$, the formalism can truly surpass some deficiencies of $f(R)$ gravity. As the $f(R,T)$ function is arbitrary, several parametric models have been proposed {\it ad hoc} in the literature and posteriorly confronted with observational data. In the present article, we use gaussian process to construct an $f(R,T)=R+f(T)$ model. To apply the gaussian process we use a series of measurements of the Hubble parameter. We then analytically obtain the functional form of the function. By construction, this form, which is novel in the literature, is well-adjusted to cosmological data. In addition, by extrapolating our reconstruction to redshift $z=0$, we were able to constrain the Hubble constant value to $H_0=69.97\pm4.13$$\rm \ km \ s^{-1} \ Mpc^{-1}$ with $5\%$ precision. Lastly, we encourage the application of the functional form herewith obtained to other current problems of observational cosmology and astrophysics, such as the rotation curves of galaxies. 


\end{abstract}  

\maketitle

\section{Introduction}\label{sec:int}

Theoretical physicists have been attempting to discover the cause of the acceleration in the expansion of the universe for some decades. Although the observational evidence for a cosmic acceleration first came around two and a half decades ago, from the detected diminishing in the brightness of distant supernovae Ia \cite{riess/1998,perlmutter/1999}, we still do not have the ultimate answer for what essentially makes the expansion of the universe to accelerate. 

The first attempt to explain the cosmic acceleration was indirectly proposed by Einstein himself and is called {\it cosmological constant} $\Lambda$. In the presence of the cosmological constant, the Einstein's field equations of General Relativity, $G_{\mu\nu}=8\pi T_{\mu\nu}-\Lambda g_{\mu\nu}$ are capable of describing the acceleration of the expansion of the universe in accordance with observations. Here, $G_{\mu\nu}$ is the Einstein tensor, $T_{\mu\nu}$ is the energy-momentum tensor and $g_{\mu\nu}$ is the metric tensor; moreover we will work with natural units. The cosmological constant is the same that does not allow the space-time to be flat in the absence of sources, as the above equation in this regime reads $R_{\mu\nu}=\Lambda g_{\mu\nu}$, with $R_{\mu\nu}$ being the Ricci tensor. $\Lambda$ is then associated with the vacuum, which intriguingly curves space. In order to adjust to the cosmological observations, $\Lambda$ must be $\sim10^{-52}$m \cite{riess/1998,perlmutter/1999,padmanabhan/2002}.

The problem rises when we calculate the theoretical value of the energy density of vacuum, which is higher than the density of $\Lambda$ by up to $120$ orders of magnitude \cite{weinberg/1989}. This spectacularly high disagreement is referred to as the {\it cosmological constant problem}. 

An alternative is to attempt to treat the cosmic acceleration with no cosmological constant, for instance, explaining such an acceleration as due to a scalar field. These are the so-called {\it quintessence models} (check, for instance, References \cite{tsujikawa/2013,wang/2000,linder/2008}), in which a scalar field governs the dynamics of the universe, and depending on the form of the potential of such a scalar field, it is possible to explain the cosmic acceleration in accordance with observational data. 

Another possibility to evade the cosmological constant problem is to consider modifications or extensions of General Theory of Relativity. In such an approach, the cosmic acceleration rises as a geometrical effect of this ``broader'' theory of gravity. The extra terms of extended gravity could also naturally induce the existence of a fluid which is capable of generating the repulsive feature necessary to accelerate the universe expansion. 

An example of extended gravity theory is the well-known $f(R)$ gravity \cite{de_felice/2010}, in which ``$f(R)$'' stands for a generic function of the Ricci scalar $R$ to substitute $R$ in the Einstein-Hilbert action. Naturally, this substitution leads to the appearance of extra terms in the resulting field equations, and those could be responsible for driving the universe expansion to accelerate, as aforementioned. In fact, it has been shown to be possible to accelerate the expansion of the universe through $f(R)$ modifications of gravity in \cite{capozziello/2005,song/2007}.

Naturally, an extended gravity theory should be well-performed in different regimes of applications; that is to say that besides explaining the cosmic acceleration with no need for the cosmological constant, it should also behave properly in the stellar and galactic regimes, in the Solar System regime etc. Some $f(R)$ gravity shortcomings are discussed in \cite{joras/2011,casado-turrion/2022}, among other references. Some other shortcomings are mentioned in \cite{harko/2011}. 

Still in \cite{harko/2011}, the $f(R,T)$ gravity theory was proposed, for which the generic function $f$ that substitutes $R$ in the Einstein-Hilbert action also depends on the trace of the energy-momentum tensor $T$. While in the $f(R)$ gravity, the dependence on higher order terms of $R$ is motivated by the possibility that gravity indeed may behave differently in cosmological scales, the $T-$dependence is motivated by the possible existence of imperfect and/or extra fluids permeating the universe. It could also be related to quantum effects \cite{harko/2011} or an effective cosmological constant ``$\Lambda(T)$''. Some technical discussion on the $T$-dependence terms of the $f(R,T)$ gravity can be seen in \cite{harko/2020}. 

The list of applications of $f(R,T)$ gravity is quite extensive and we mention a few of them in the following. Traversable wormhole solutions can be seen in \cite{sahoo/2018,moraes/2017,moraes/2019,moraes/2017b}. The Tolman-Oppenheimer-Volkoff-like equation was first derived and solved in \cite{moraes/2016}. In \cite{alves/2016}, the extra polarization states of gravitational waves were calculated for the $f(R,T)$ gravity. The effects of $f(R,T)$ gravity on gravitational lensing were calculated in \cite{alhamzawi/2016}. The Solar System consequences of the theory can be seen in \cite{shabani/2014}. 

The $f(R,T)$ function is generic, although some constraints are commonly put to it, for instance, from the energy conditions \cite{alvarenga/2013}. In the $f(R,T)$ gravity, in a first analysis, the energy-momentum tensor does not conserve, although some ``conservative models'' were presented in the literature. For instance, in a cosmological perspective, it was shown in \cite{velten/2017} that a conservative model like $f(R,T)=R+\alpha\sqrt{T}$, with $\alpha$ being a free parameter, is capable of describing  the cosmological observational data in great accordance. Moreover, the hydrostatic equilibrium configurations of neutron stars were presented for a conservative $f(R,T)$ model in \cite{dos_santos_jr/2019}, and the results are capable of predicting the existence of massive pulsars, such as those reported in \cite{antoniadis/2013,nice/2005,linares/2018}, differently from the results obtained from a non-conservative version of the theory \cite{moraes/2016}, which cannot reach such high masses.  

There are several other alternative gravity theories that underline cosmological models that accelerate the universe expansion, such as Gauss-Bonnet gravity \cite{dehghani/2004}, extra-dimensional models \cite{gu/2002} and $f(\mathcal{T})$ gravity \cite{cai/2016}, for which $\mathcal{T}$ stands for the torsion scalar. 

It is worth remarking that alternative gravity theories are also motivated by the lack of dark matter particle detection \cite{schumann/2019,boehm/2004,marrodan_undagoitia/2016,munoz/2004,akerib/2014}. There are several astrophysical evidences pointing to dark matter existence, such as those reported in References \cite{chen/2022,iocco/2015,mclaughlin/1999,conselice/2021}, among others. On the other hand, there are several dark matter particles modelling in the present literature \cite{bertone/2005,jungman/1996,feng/2010,bernal/2017,servant/2003,bringmann/2009}, but with no experimental counterpart, as just mentioned. This leads to the possibility that dark matter is simply a gravitational effect of an extended theory of gravity \cite{choudhury/2016,capozziello/2013,capozziello/2007}.

Despite the problems regarding the ``dark sector'' of the universe in standard cosmology, namely dark energy and dark matter, there is also another observational issue nowadays, which is referred to as the {\it Hubble tension}. The Hubble tension is a discrepancy between measures of the universe expansion rate obtained by different observational methods \cite{lopez-corredoira/2022,krishnan/2021,thakur/2023,schoneberg/2019,mortsell/2022}. Modified gravity has also been invoked to treat the Hubble tension \cite{sivaram/2022,abadi/2021,shimon/2022,odintsov/2021}.

In the aforementioned $f(\mathcal{T})$ gravity, $f$ is also a generic function of the argument. In \cite{cai/2020}, the gaussian processes (GPs) and Hubble function data were applied to $f(\mathcal{T})$ cosmology to construct the $f(\mathcal{T})$ functional form for the first time. The GP will be explained below. For now, it is worth mentioning that it is a powerful tool one can use to construct the behavior of a function directly from a data set \cite{seikel/2012}. 

In the present article we will apply the GP using Hubble parameter $H(z)$ data in order to construct the $f(R,T)$ functional form. The article is organized as follows. In Section \ref{sec:frt} we present the $f(R,T)$ gravity, its field equations and we obtain its Friedmann-like equations. In Section \ref{sec:recfrt} we present the GP, and numerically and analytically reconstruct the $f(R,T)$ function. We discuss our results and present our concluding remarks in Section \ref{sec:dis}.

\section{The $f(R,T)$ gravity}\label{sec:frt}

Proposed by Harko and collaborators in Reference \cite{harko/2011}, the $f(R,T)$ theory of gravity starts from the action

\begin{equation}\label{frt1}
S=\int d^4x\sqrt{-g}\left[\frac{f(R,T)}{16\pi}+L\right].    
\end{equation}
In the above equation, $g$ is the metric determinant and $L$ is the matter lagrangian density. Moreover we will assume natural units, such that $G=c=1$.

The $f(R,T)$ gravity field equations are obtained from the variation of the above action with respect to the metric $g_{\mu\nu}$ and yield

\begin{equation}\label{frt2}
(R_{\mu\nu}+g_{\mu\nu}\Box-\nabla_\mu\nabla_\nu)f_R-\frac{fg_{\mu\nu}}{2}=8\pi T_{\mu\nu}+f_T(T_{\mu\nu}-Lg_{\mu\nu}),   
\end{equation}
with $R_{\mu\nu}$ being the Ricci tensor, $f_R\equiv\partial f/\partial R$,

\begin{equation}\label{frt3}
T_{\mu\nu}=\frac{-2}{\sqrt{-g}}\frac{\partial(\sqrt{-g}L)}{\partial g^{\mu\nu}}  
\end{equation}
the energy-momentum tensor and $f_T\equiv\partial f/\partial T$. 

In order to proceed and with the purpose of particularly investigating the $T-$dependence of the theory and its consequences, we will from now on assume $f(R,T)=R+2f(T)$. This assumption has been made in several works, such as \cite{moraes/2017,azizi/2013,alvarenga/2013,shabani/2018,moraes/2019,moraes/2017b,godani/2022,velten/2017,sardar/2023}. We consider that the Universe can be modeled as a perfect fluid consisting entirely of matter, with no other forms of energy or pressure present. The above assumptions allow us to express

\begin{eqnarray}\label{feqs}
R_{\mu\nu}-\frac{1}{2} R g_{\mu\nu}=8\pi T_{\mu\nu}+2f_T T_{\mu\nu}+f(T)g_{\mu\nu}.
\end{eqnarray}

In order to establish the background cosmological framework, we assume the principle of homogeneity and isotropy of the Universe by adopting the Friedmann-Lemaître-Robertson-Walker metric

\begin{equation}
    ds^2=dt^2-a^2(t)\delta_{ij}dx^idx^j
\end{equation}
where $a(t)$ is the scale factor. Applying the metric into Equation (\ref{feqs}), we write the Friedmann equations of $f(R,T)$ gravity as follows:

\begin{eqnarray}
H^2&=&\frac{8\pi\rho_m}{3}+\frac{2f_T}{3}\rho_m-\frac{f(T)}{3};\\\label{fridfrt}
\dot H +H^2&=&\frac{\ddot a}{a}=-\frac{4\pi}{3}\rho_m-\frac{1}{3}\left[f_T\rho_m+f(T)\right],
\end{eqnarray}
in which $H\equiv \dot a/a$ is the Hubble parameter, $\rho_m$ is the energy density of matter and dots represent derivative with respect to time $t$. The application of the GPs method, which will be presented in the next section, will rely on these equations as a fundamental component. In particular, if we consider a universe comprised entirely of matter and set the function $f(T)$ as $f(T)=2T$, with $p=0$, then $T=-\rho$, and the resulting theory can be regarded as a cosmological model with an effective cosmological constant that is proportional to the Hubble parameter.

\section{Reconstruction of $f(R,T)$ through Gaussian Process}\label{sec:recfrt}

\subsection{Gaussian Process}

The GP can be employed as a non-linear regression approach for reconstructing a function using observational data, without assuming a parametric model. These processes can be seen as a collection of random variables, where each finite set is associated with a Gaussian distribution \cite{williams2006gaussian, kocijan2016modelling, seikel/2012}. 

%A Gaussian Process is a generalization of a Gaussian distribution, describing a distribution over functions that is completely specified by its mean and covariance functions. In contrast, a Gaussian distribution is related to the distribution of a random variable.

A GP can be considered as a generalization of a Gaussian distribution. While a Gaussian distribution is related to the distribution of a random variable, a GP describes a distribution over functions in such a way that it is completely specified through its mean and covariance functions. When a function $f$ is reconstructed via GPs, the value of $f$ computed at a point $x$ is a random variable with mean $\mu(x)$ and variance $\text{Var}(x)$. This value is dependent on the value of the same function evaluated at $\tilde{x}$, particularly when these points are in close proximity to each other. These values are related through a covariance function $\text{cov}\left( f(x),f(\tilde{x}) \right) = k(x,\tilde{x})$, also known as a kernel in the computational context. Therefore, the distribution of functions can be expressed by the following equations \cite{williams2006gaussian, seikel/2012}:

\begin{eqnarray}
\mu(x) &=& \mathbb{E}[f(x)]\,,\\
k(x,\tilde{x}) &=& \mathbb{E}[(f(x)-\mu(x))(f(\tilde{x})-\mu(\tilde{x}))]\,,\\  
\text{Var}(x) &=& k(x,x)\,,
\end{eqnarray}
where the GP is expressed as:

\begin{equation}
f(x) \sim \mathcal{GP}\left( \mu(x), k(x,\tilde{x})
\right) \;.
\end{equation}

When dealing with regression problems \cite{kocijan2016modelling}, it is common to approximate the nonlinear function that represents the observed data and the measured points (e.g., Hubble parameter $H(z)$ at a specific redshift $z$) using a set of basis functions. These functions have parameters that are optimized during the regression process, and the kernel function is derived from them. Although there are several types of kernel functions used in different computational approaches, this work focuses on the squared exponential covariance function:

\begin{equation}
k(x,\tilde{x}) =
\sigma_f^2 \exp\left( -\frac{(x - \tilde{x})^2}{2\ell^2} \right) \;.
\end{equation}

The squared exponential kernel function is a covariance between the output data, expressed in terms of the input data, used to approximate a non-linear function that represents observational data and the points at which they were measured, such as the Hubble parameter $H(z)$ at a specific redshift $z$, during regression problems \cite{kocijan2016modelling}. The covariance approaches 1 when the input variables are close and decreases as the distance between them increases. Additionally, this kernel function is infinitely differentiable, making it useful for reconstructing derivatives of a function. The hyperparameters $\sigma_f$ and $\ell$ characterize the Gaussianity of the kernel function. $\sigma_f$ represents the distance one must travel on the $x$ axis to compute a significant change in $f(x)$, while $\ell$ represents a change in the $y$ axis. This information can be found in \cite{williams2006gaussian, seikel/2012}.

\subsection{Data and numerical reconstruction}

To reconstruct the $f(R,T)$ function, the first step is to utilize a series of measurements of the Hubble parameter, $H(z)$, to apply GPs and deduce its evolution. This process will enable the reconstruction of a range of $f(R,T)$ functions using the Friedmann equation, which can then be compared with the current concordance cosmological model ($\Lambda$ Cold Dark Matter model).

\begin{figure}[h!]
 \includegraphics[width=\columnwidth]{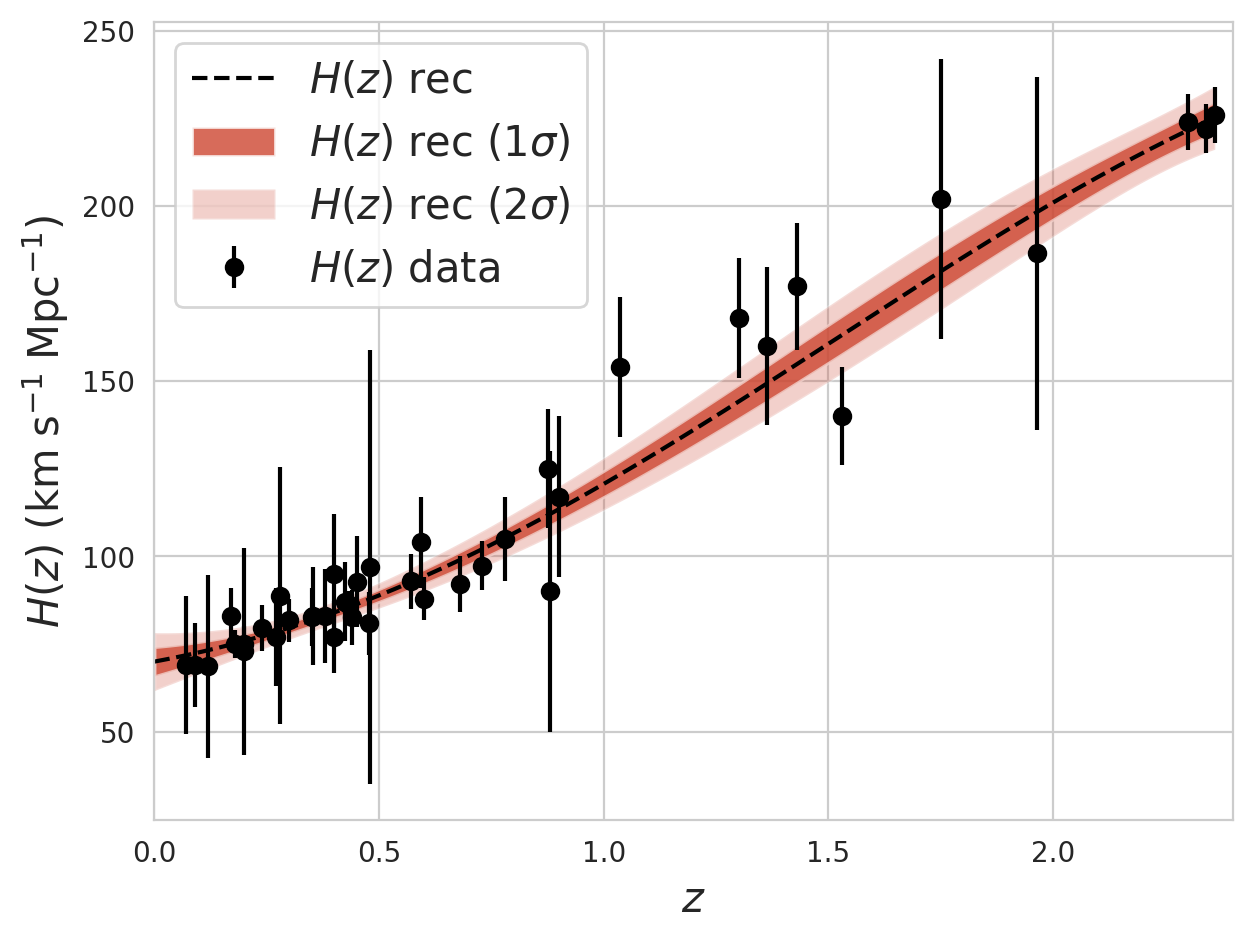}
 \caption{The reconstructed evolution of $H(z)$ from the data using Gaussian Process. The $1$ and $2$ sigma regions of statistical confidence are plotted as the orange-shaded area.}
 \label{hzrec}
\end{figure}

We reconstruct the $H(z)$ function using a dataset compiled in \cite{jesus2018bayesian}, which included 41 measurements of $H(z)$ obtained from two distinct methods -- displayed in Table \ref{hzdata} below. The Cosmic Chronometers method measures the age difference between pairs of ancient spiral galaxies that formed at similar epochs and redshifts, while the method based on the position of the peak of the Baryonic Acoustic Oscillations provides a standard ruler in the radial direction by measuring clusters of galaxies \cite{padilla2021cosmological, jimenez2002constraining,li2021testing}. We processed these observational data using a publicly available Python algorithm  \verb|GAPP| developed by Seikel et al. \cite{seikel/2012}. This enabled us to reconstruct $H(z)$ with statistical confidence regions at $1\sigma$ and $2\sigma$ levels in Figure \ref{hzrec}. The reconstructed function $H(z)$ will be used to reconstruct the $f(R,T)$ function, as described above.

\begin{table*}%[p!]
\begin{center}
\begin{tabular}[t]{c c c c}
\hline
\textbf{$z$} & \textbf{$H(z)$} & \textbf{$\sigma_H$} & \text{Reference}\\
\hline
0.070 & 69 & 19.6 & \cite{zhang2014four}\\
0.090 & 69 & 12 & \cite{simon2005constraints}\\
0.120 & 68.6 & 26.2 & \cite{zhang2014four}\\
0.170 & 83 & 8 & \cite{simon2005constraints}\\
0.179 & 75 & 4 & \cite{moresco2012improved}\\
0.199 & 75 & 5 & \cite{moresco2012improved}\\
0.200 & 72.9 & 29.6 & \cite{zhang2014four}\\
0.240 & 79.69 & 6.65 & \cite{gaztanaga2009clustering}\\
0.270 & 77 & 14 & \cite{simon2005constraints}\\
0.280 & 88.8 & 36.6 & \cite{zhang2014four}\\
0.300 & 81.7 & 6.22 & \cite{oka2014simultaneous}\\
0.350 & 82.7 & 8.4 & \cite{chuang2013modelling}\\
0.352 & 83 & 14 & \cite{moresco2012improved}\\
0.3802 & 83 & 13.5 & \cite{moresco20166}\\
0.400 & 95 & 17 & \cite{simon2005constraints}\\
0.4004 & 77 & 10.2 & \cite{moresco20166}\\
0.4247 & 87.1 & 11.2 & \cite{moresco20166}\\
0.430 & 86.45 & 3.68 & \cite{gaztanaga2009clustering}\\
0.440 & 82.6 & 7.8 & \cite{blake2012wigglez}\\
0.4497 & 92.8 & 12.9 & \cite{moresco20166}\\
0.4783 & 80.9 & 9 & \cite{moresco20166}\\
\hline
%\end{tabular}
\end{tabular}
%\begin{tabular}[t]{ll}
\begin{tabular}[t]{c c c c}
\hline
\textbf{$z$} & \textbf{$H(z)$} & \textbf{$\sigma_H$} & \text{Reference}\\
\hline
0.480 & 97 & 62 & \cite{stern2010cosmic}\\
0.570 & 92.900 & 7.855 & \cite{anderson2014clustering}\\
0.593 & 104 & 13 & \cite{moresco2012improved}\\
0.6 & 87.9 & 6.1 & \cite{blake2012wigglez}\\
0.680 & 92 & 8 & \cite{moresco2012improved}\\
0.73 & 97.3 & 7.0 & \cite{blake2012wigglez}\\
0.781 & 105 & 12 & \cite{moresco2012improved}\\
0.875 & 125 & 17 & \cite{moresco2012improved}\\
0.880 & 90 & 40 & \cite{stern2010cosmic}\\
0.900 & 117 & 23 & \cite{simon2005constraints}\\
1.037 & 154 & 20 & \cite{moresco2012improved}\\
1.300 & 168 & 17 & \cite{simon2005constraints}\\
1.363 & 160 & 22.6 & \cite{moresco2015raising}\\
1.430 & 177 & 18 & \cite{simon2005constraints}\\
1.530 & 140 & 14 & \cite{simon2005constraints}\\
1.750 & 202 & 40 & \cite{simon2005constraints}\\
1.965 & 186.5 & 50.4 & \cite{moresco2015raising}\\
2.300 & 224 & 8 & \cite{delubac2013baryon}\\
2.34 & 222 & 7 & \cite{delubac2015baryon}\\
2.36 & 226 & 8 & \cite{font2014quasar}\\
\hline

%\end{tabular}

\end{tabular}
\end{center}
\caption{41 measures of $H(z)$ used for the reconstruction of $f(R,T)$. \label{hzdata}}
\end{table*}

In our reconstruction of the $f(R,T)=R+2f(T)$ function, we adopted a methodology similar to that proposed in \cite{briffa2021constraining} for the case of teleparallel gravity. Given that the dynamics of $f(T)$ are governed by the Friedmann equation via Eq.(\ref{fridfrt}), we express all relevant quantities in terms of redshift and leveraged our reconstruction of $H(z)$ to obtain the desired curve. To approximate the function for small redshift differences $\Delta z$, we employ the following expression:
\begin{equation}
f_{T} \equiv \frac{df(T)}{dT} = \frac{df/dz}{dT/dz}=\frac{f'}{T'} ~, 
\end{equation}
\noindent where $T^\prime = -(9H_0^2\Omega_{m0}/8\pi)(1+z)^2$. Then, it is assumed that $f^\prime$ is given by:
\begin{equation}\label{eq:fprime}
f^{\prime}(z) \approx \frac{f(z+\Delta z) - f(z)} {\Delta z} ~.
\end{equation}
\noindent In this way, it is possible to relate the values of $f$ at $z_{i+1}$ and $z_i$. By applying this relation to the Friedmann equation, we obtain:
\begin{eqnarray}\nonumber
    f(z_{i+1})&=&f(z_i)-\frac{9}{2}\frac{\left(z_{i+1}-z_i\right)}{\left(1+z_i\right)}\bigg[H^2\\
    &-&H_0^2\Omega_{m0}\left(1+z_i\right)^3+\frac{f(z_i)}{3}\bigg].
\end{eqnarray}

\begin{figure}[h!]
 \includegraphics[width=\columnwidth]{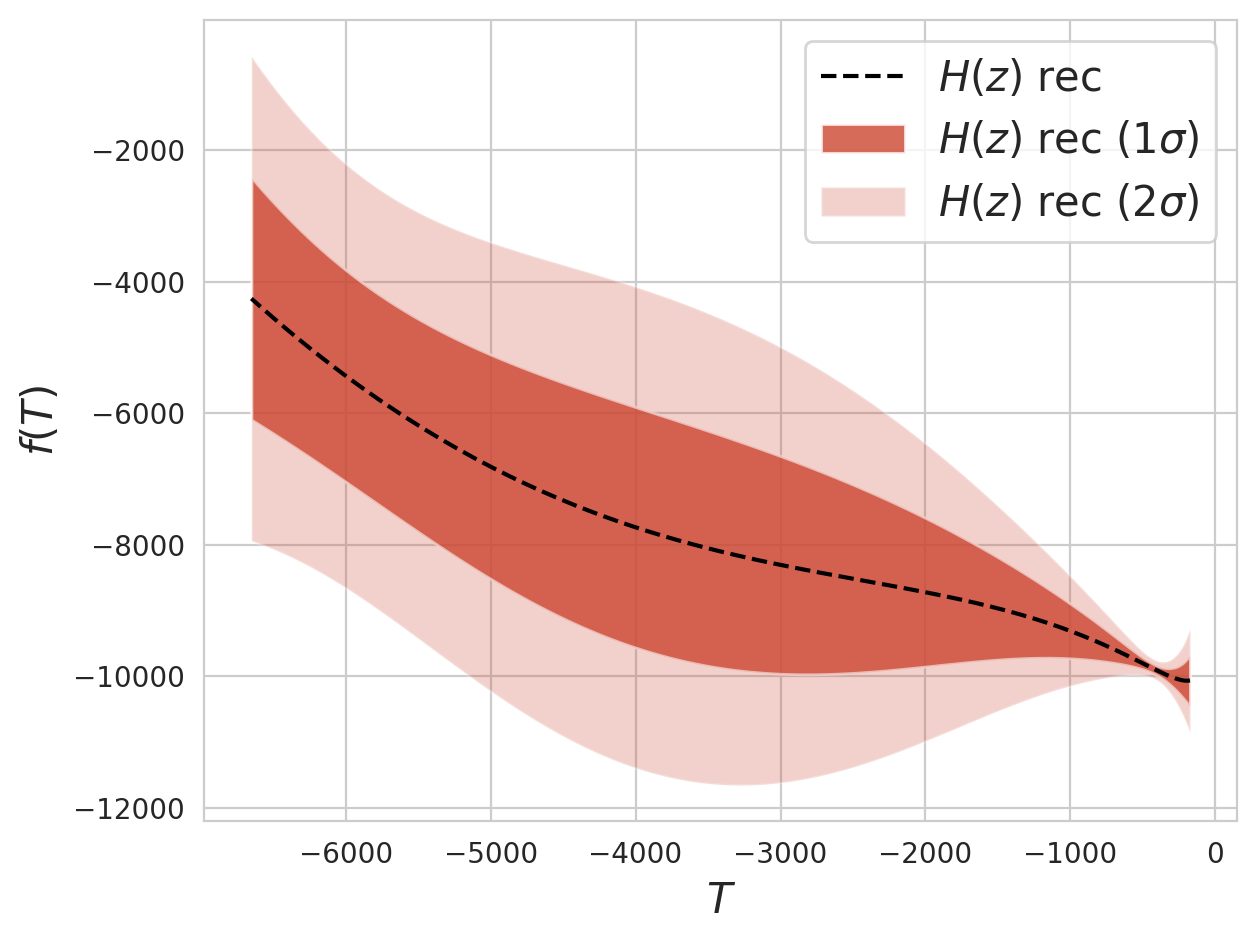}
 \caption{The reconstructed $f(T)$ curve plotted alongside its regions of statistical confidence.}
 \label{ftrec}
\end{figure}

To apply GPs, an initial condition needs to be imposed to solve the above recurrence relation. Therefore, we compute the Friedmann equation for $z=0$ by assuming $f_T\approx0$, which results in:
\begin{equation}
    f(z=0)=3H_0^2\left(\Omega_{m0}-1\right), 
\end{equation}
\noindent with $\Omega_{m0}=0.315\pm0.007$ from the last Planck data release \cite{aghanim2020planck}. Thus, given the relationship between $T$ and $H(z)$, it is possible to reconstruct the $f(T)$ function and consequently the $f(R,T)$ function. The results are shown in Figure \ref{ftrec}, where the statistical confidence regions are plotted.

\subsection{Analytical fit of the reconstructed function}

To gain deeper insight into the behavior of the reconstructed function $f(T)$, it is helpful to express it analytically through polynomial fitting. This approach serves as a cross-check to validate our findings and has been employed in other studies focused on model-independent reconstructions \cite{cai/2020, briffa2021constraining, ren2021data}. 

Through the use of 200 reconstructed $f(T)$ points obtained via GP, we analysed certain analytical forms: the quadratic polynomial $f(T)$ model, $f(T)=\alpha T^2+\beta$; the cubic polynomial $f(T)$ model, $f(T)=\alpha T^3+\beta T^2+\gamma$; and the hyperbolic tangent polynomial $f(T)=\alpha T^2+A\tanh\left[\lambda\left( T+T_0\right)\right]+\beta T+ \gamma$ (which we will, from now on, refer to as HTP model), in which $\alpha, \beta, \gamma, A, \lambda$ and $T_0$ are constants.

To select the model with the best fitting behavior, first we plug every considered analytic $f(T)$ curve into the Friedmann equation to find the evolution of the Hubble parameter $H(z)$, then we calculate the $\chi^2$ statistics as follows:
\begin{equation}
\chi^2_{H(z)}=\sum^{41}_{i=1}\frac{\left(H_{obs, i}-H_{fit, i}\right)^2}{\sigma^2_{H_{obs,i}}},
\end{equation}
\noindent with $H_{obs}$ being the Hubble parameter measurements, $\sigma_{H_{obs}}$ the related errors and $H_{fit}$ the values provided by the analytic fitting. 

From the $\ln\mathcal{L}\propto -0.5\chi^2$ relation, we use Bayesian Information Criterion (BIC) \cite{schwarz1978estimating}, which is defined as the following:
\begin{equation}
    \rm BIC = -2\ln\mathcal{L}_{max} +2 \rm p\ln(\rm n),
\end{equation}
\noindent where $\mathcal{L}$ is the likelihood of the data given the model, $\rm p$ is the number of parameters in the model and $\rm n$ is the number of data points, to evaluate the goodness of the fit for various models. A lower BIC value indicates a better fit with fewer parameters. In Table \ref{bicc}, we present the BIC values for all the models studied in this work. The results show that the HTP model has the lowest BIC value, indicating the best fit among the tested models.

\begin{table}[h!]
\begin{center}
\begin{tabular}[t]{c | c | c}
\hline
\text{Model} & \text{$\rm BIC$} & \text{$\rm \Delta BIC$} \\
\hline
$\Lambda CDM$ & 37.89 & 9.61\\
quadratic polynomial $f(T)$ model & 34.63 & 6.35\\
cubic polynomial $f(T)$ model & 42.35 & 14.7\\
%quartic polynomial $f(T)$ model & 48.28 & 20\\
hyperbolic tangent polynomial $f(T)$ model & 28.28 & 0\\
\hline
%\end{tabular}
\end{tabular}
\end{center}
\caption{The BIC and $\Delta \rm BIC$ values for the $f(T)$ models considered in the present work. \label{bicc}}
\end{table}

To compare the models and assess their relative strength of evidence, we calculate the difference between the BIC value for each model and the minimum BIC value, denoted as $\Delta \rm BIC = \rm BIC - \rm BIC_{min}$. A difference of less than 2 suggests weak evidence, while a difference of 6 or more indicates very strong evidence in favor of the better-performing model. Therefore, a larger $\Delta \rm BIC$ value suggests that one model is significantly more likely to be the true model than another. 

\begin{figure}[h!]
 \includegraphics[width=\columnwidth]{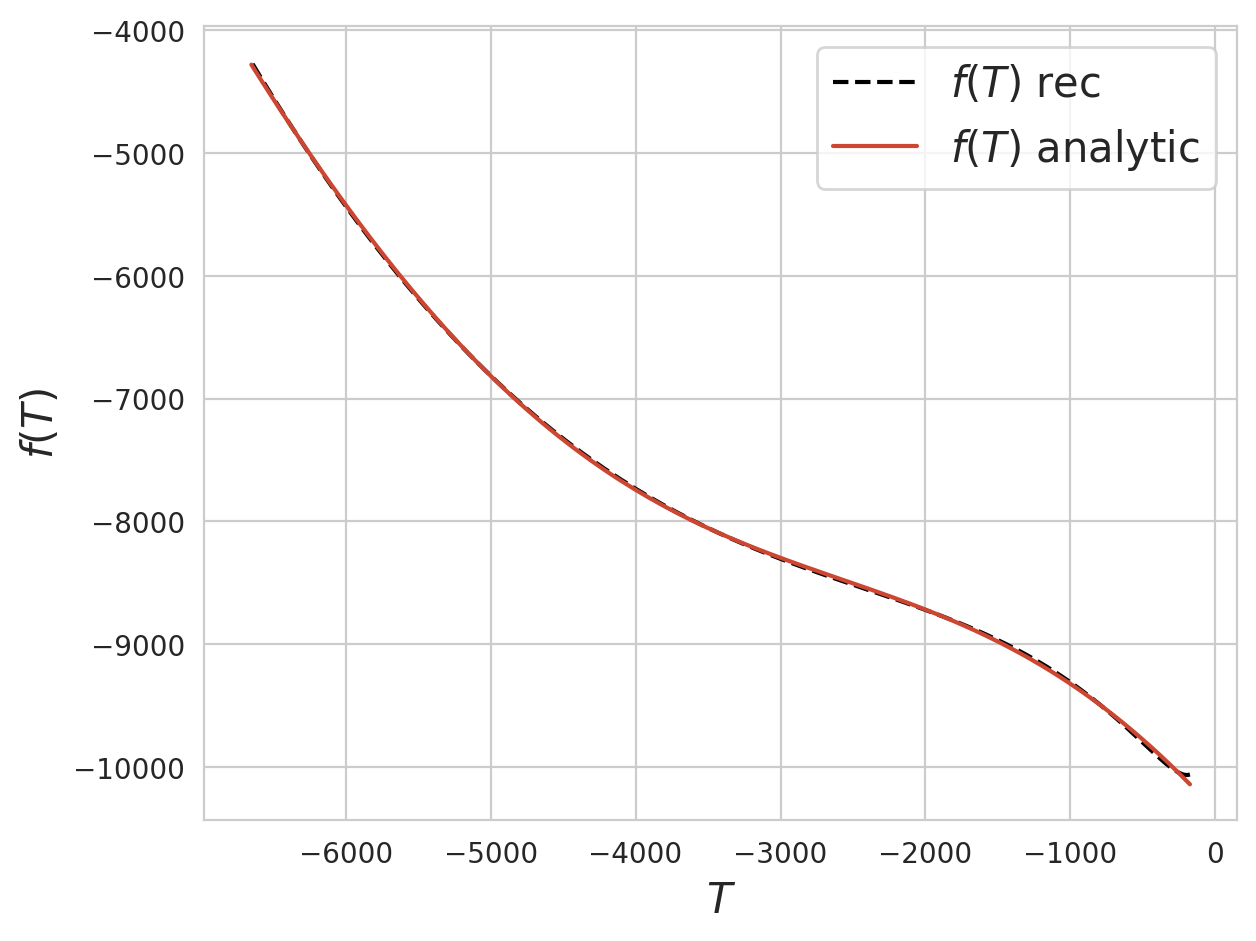}
 \caption{The analytic fit of the reconstructed $f(T)$ function (black-dashed line) from the hyperbolic tangent polynomial model (orange-solid line).}
 \label{analitic}
\end{figure}

Our analysis shows that the HTP model is the best fit for the data, as it has the lowest BIC value and a significant difference in $\Delta \rm BIC$ compared to the other models, that is, the HTP model is the preferred model for describing the behavior of the reconstructed $f(T)$ curve, with the following values for its parameters: $\alpha = -1.83\times 10^{-5}$, $A=-1.05\times 10^4$, $\lambda=-2.39\times 10^3$, $T_0=2.58\times 10^3$, $\beta=-2.99$ and $\gamma=-1.61\times 10^4$. In Figure \ref{analitic} we plot the preferred model in comparison to the GP reconstructed one.

Now, to further validate the HTP model here obtained, we compare the evolution of $H(z)$ using the Friedmann equation with that of the $\Lambda \rm CDM$ model in Figure \ref{comp} below. 

Figure \ref{comp} shows the $H(z)$ function for the preferred analytic model, as well as the $\Lambda \rm CDM$ model using the latest Planck data release \cite{aghanim2020planck}, and the GPs reconstructed relations for the same parameter. The comparison reveals that the $H(z)$ curves are in strong agreement for redshifts approximately less than $1.7$. However, for larger redshift values, the $\Lambda \rm CDM$ curve shows a discrepant behavior, which may be attributed to the lack of $H(z)$ measurements in this range. This provides further support for the HTP model as a better fit for the data. The Friedmann equation comparison serves as another useful cross-check to validate our findings and reinforces the effectiveness of the HTP model for describing the behavior of the reconstructed $f(T)$ curve.

\begin{figure}[h!]
 \includegraphics[width=\columnwidth]{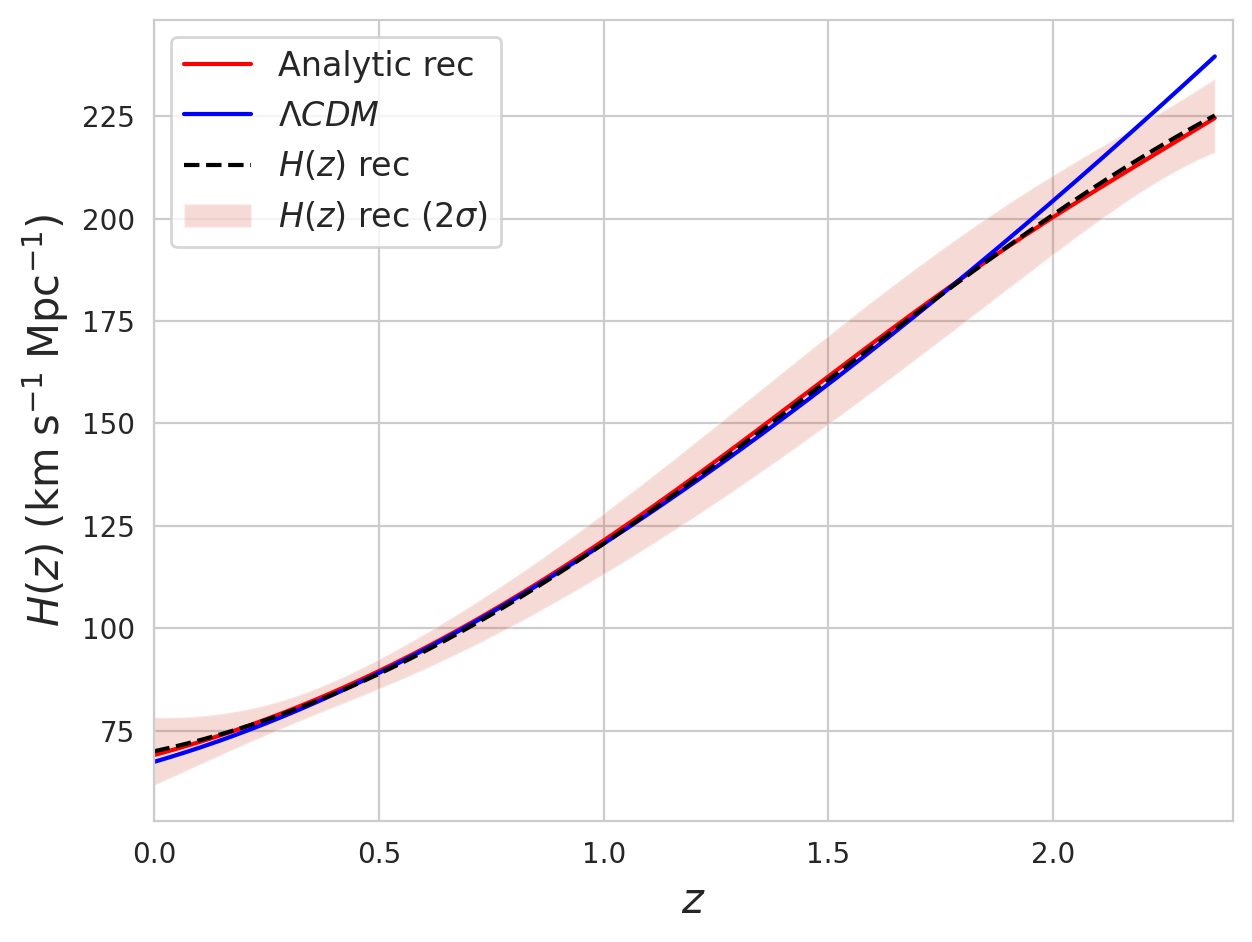}
 \caption{Comparison between the hyperbolic tangent $f(T)$ model (red-solid line) and the $\Lambda \rm CDM$ cosmological concordance model (blue-solid line). The reconstructed evolution of $H(z)$ is plotted as a black-dashed line alongside its $2\sigma$ uncertainties as a shaded area.}
 \label{comp}
\end{figure}

\begin{figure}[h!]
 \includegraphics[width=\columnwidth]{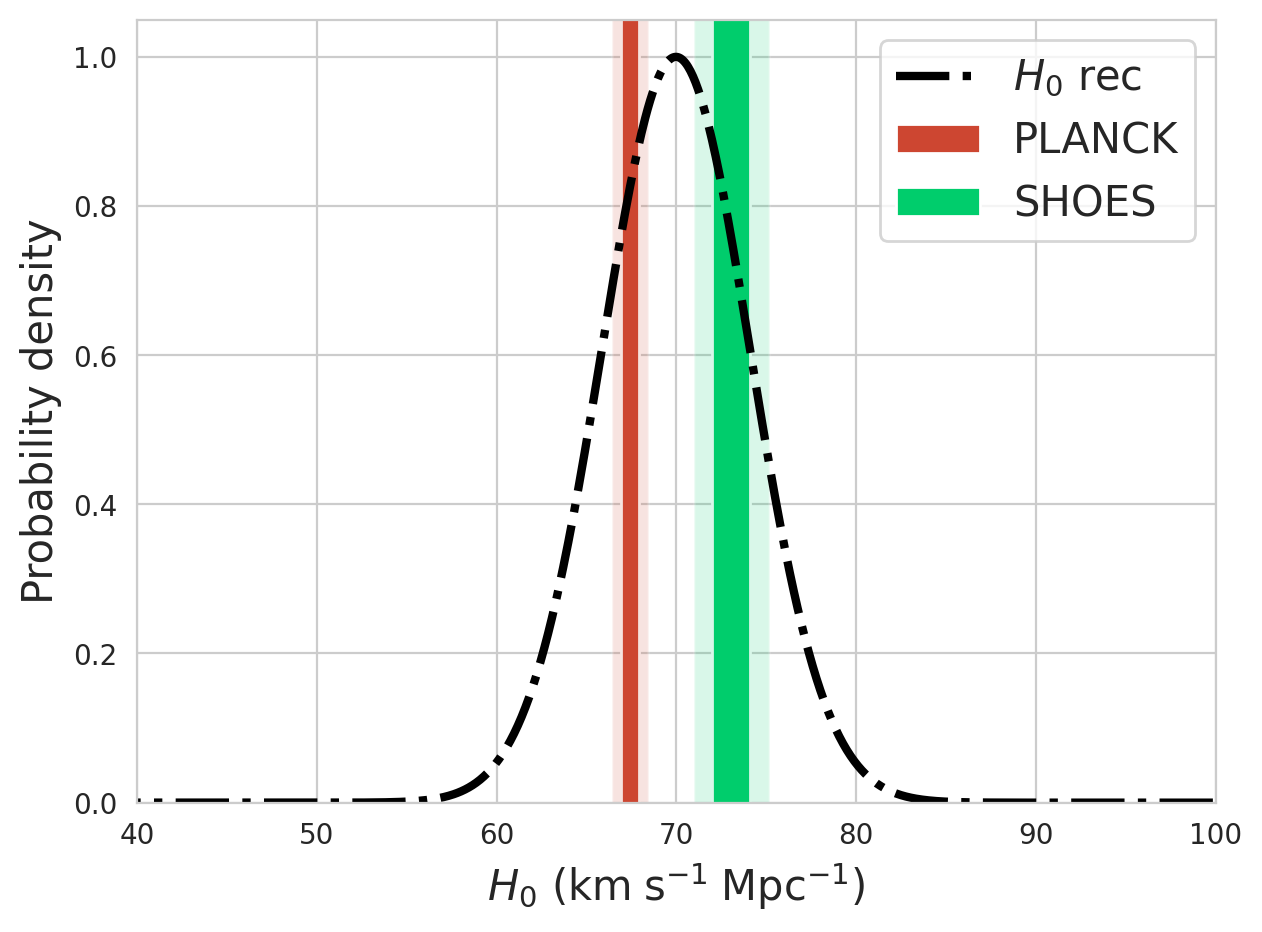}
 \caption{The reconstructed Hubble constant found in this study compared to the values of the last Planck data release and the value provided by the SH0ES collaboration.}
 \label{h0}
\end{figure}

We also stress that by using the HTP model into the Friedmann equation and extrapolating to $z=0$ we can constrain the Hubble parameter to $H_0 = 69.97\pm4.13$$\rm \ km \ s^{-1} \ Mpc^{-1}$, which is consistent with previous measurements \cite{aghanim2020planck,riess2022comprehensive}. Figure \ref{h0} displays our result alongside the last Planck data release and the SH0ES team. 

\section{Discussions}\label{sec:dis}

Many alternatives to General Relativity have been proposed with the main purpose of evading the cosmological constant problem. Other problems, such as the dark matter problem and the Hubble tension also motivate the search for broader theories of gravity. 

Among the alternatives that we see in the present literature, the $f(R,T)$ gravity has emerged as one of the most optimistic possibilities. Besides the $f(R,T)$ gravity applications that we mentioned in the Introduction, we should add the following. The Palatini formulation of $f(R,T)$ gravity and its cosmological consequences can be seen in \cite{wu/2018}.  The study of scalar cosmological perturbations was made in \cite{alvarenga/2013b}. Different cosmological models, such as Chaplygin gas and quintessence models, were reconstructed in the framework of $f(R,T)$ gravity in \cite{jamil/2012}. $f(R,T)$ cosmological solutions were investigated through the phase-space analysis in \cite{shabani/2013}. A particular model to describe inflation in the $f(R,T)$ gravity was presented in \cite{bhattacharjee/2020}. Further wormhole solutions in $f(R,T)$ gravity can be seen in \cite{elizalde/2019}. The idea of ``unimodular gravity'' was extended to $f(R,T)$ theory in \cite{rajabi/2017}. Finally, in \cite{moraes/2016b}, $T$ was taken as the energy-momentum tensor of a scalar field, and for a given scalar field potential, the obtained cosmological model was able to describe the complete history of the universe evolution, including a graceful exit from the inflationary to the radiation-dominated era.

In the present article, we have, for the first time in the literature, implemented the Gaussian Process to the $f(R,T)$ gravity framework. Through analysis of 41 measurements of the Hubble parameter $H(z)$, we were able to reconstruct both numerically and analytically the $f(T)$ function within $f(R,T)=R+f(T)$, by arriving in a novel $f(R,T)$ model: the HTP model. Using the Bayesian Information Criterion, our analysis provides strong evidence in support of the HTP model when compared to the current cosmological concordance model. In the current scenario of extended gravity theories, this raises as a good alternative to evade the cosmological constant problem. 

The introduction of this new model opens up unexplored avenues for testing against further cosmological and astrophysical data. For instance, in the Introduction we have mentioned that not only the cosmological constant problem needs deep attention in standard cosmology framework, but also the absence of dark matter detection. It has, indeed, been shown that the dark matter gravitational effects in the rotation curves of galaxies can be simply predicted in the weak-field regime of broader gravity formalisms \cite{mak/2004,harko/2010}. The next step would be to apply the HTP model here obtained to the galactic dynamics regime to check if it is also capable of describing dark matter effects.

A third observational issue that needs attention in standard cosmology is the Hubble tension (also mentioned in the Introduction). Could it also be analysed under extended gravity? Extrapolation of our analysis to redshift $z=0$ yields a value of $H_0=69.96\pm4.13$$\rm \ km \ s^{-1} \ Mpc^{-1}$ for the Hubble constant, consistent with other works \cite{freedman2020calibration, colas2020efficient} at the $1\sigma$ statistical confidence level. It is remarkable that our value for $H_0$ fits both Planck data and SH0ES collaboration value, which could be understood as an alleviation of the Hubble tension under the HTP model. Of course, it is important to consider the uncertainties that arise from our results, which may be attributed to the need for additional $H(z)$ measurements at high redshifts. A more precise determination of $H_0$ has the potential to provide valuable insights into the tension and maybe offer a better understanding of the underlying gravity theory being considered. 

\section*{Acknowledgements}

JAS Fortunato thanks FAPES for financial support. JG de Lima Júnior thanks CAPES for financial support.

\bibliography{Bib.bib}

\end{document}